\documentclass[aps,amsmath,amssymb]{revtex4}
\usepackage{psfig}
\usepackage{graphicx}

\begin{document}

\title{\bf Search of $D^{*}_{sJ}$ mesons in $B$ meson decays}
\date{\today}
\author{\bf  Chuan-Hung Chen\footnote{Email: chchen@phys.sinica.edu.tw}
and Hsiang-nan Li\footnote{Email: hnli@phys.sinica.edu.tw}}

\vskip1.0cm

\affiliation{Institute of Physics, Academia Sinica, Taipei, Taiwan
115, Republic of China}

\affiliation{Department of Physics, National Cheng-Kung
University, Tainan, Taiwan 701, Republic of China}

\begin{abstract}
We propose that the search of the $B\to D^{*}_{sJ}M$ decays,
$M=D$, $\pi$ and $K$, can discriminate the different theoretical
postulations for the nature of the recently observed $D^{*}_{sJ}$
mesons. The ratio of the branching ratios $B(B\to
D^{*}_{sJ}M)/B(B\to D^{(*)}_{s}M)\approx 1$ (0.1) supports that
the $D^{*}_{sJ}$ mesons are quark-antiquark (multi-quark) bound
states. The Belle measurement of the $B\to D^{*}_{sJ}D$ branching
ratios seems to indicate an unconventional picture.
\end{abstract}
\maketitle

BaBar collaboration observed a narrow state with $J^P=0^+$,
denoted by $D^{*}_{sJ}(2317)$, from the $D^{+}_{s} \pi^{0}$
invariant mass distribution \cite{Babar1}, whose mass was
determined to be $2317.6\pm 1.3$ MeV, and whose width is
consistent with the experimental resolution, being less than $10$
MeV. This observation has been confirmed by CLEO, and another new
state $D^{*}_{sJ}(2463)$ with $J^P=1^+$ was found in the
$D^{*+}_{s} \pi^0$ channel with the mass splitting $351.6\pm 1.7
\pm 1.0$ MeV from the ordinary vector meson $D^{*}_{s}$ and with
the width being less than $7$ MeV \cite{CLEO}. It is then an
urgent subject to understand the nature of these newly observed
states, and many theoretical speculations have appeared in the
literature. In this paper we shall propose an experimental strategy,
which can make a substantial contribution to this subject.

The measured masses and widths of the new states
do not match the predictions from typical potential
models. For example, the mass and width of the scalar
$D^{*}_{sJ}(2317)$ meson were expected to be around $2.48$ GeV and
$160$ MeV \cite{PM}, respectively. It has been shown that the
masses and widths of the $D_s$ system can not be explained
simultaneously in the potential model \cite{CJ}.
To resolve the discrepancy, either the theoretical models need to
be modified, or the new mesons are unconventional bound states.
For the former, a unitarized quark model has been adopted, which
includes the coupling of the scalar meson to an OZI-allowed
two-meson channel \cite{BR}. A low-mass scalar $D_s$ meson as a
quark-antiquark state could be obtained. For the latter, the
$D^{*}_{sJ}(2317)$ meson has been interpreted as a $DK$ molecule
\cite{BCL}, a $D_s\pi$ molecule \cite{Szczepaniak}, a four-quark
state \cite{YH}, and a mixing of the conventional state and the
four-quark state \cite{BPP}. However, it was argued that the
charm-strange, and even bottom-strange, four-quark states could
not be bound \cite{L}. A lattice study in the static limit, which
predicts a larger mass for the scalar $D_s$ meson as a
quark-antiquark state, supports the multi-quark postulation
\cite{Bali}. The larger scalar mass in the quark-antiquark picture
has been confirmed by a sum-rule analysis \cite{YBD}.

Considering the above series of claims and counter claims, it is
worthwhile to look for alternative theoretical and experimental
viewpoints, which may help to clarify the controversy. For
example, it has been claimed that the existence of a new $I=0$
``$D\bar D$ bound state" with a mass less than 3660 MeV would
support the four-quark picture \cite{Nu}. Whether the $D^{*}_{sJ}$
meson radiative transition is consistent with the branching ratios
of the conventional $D_{s0}^*$ and $D_{s1}$ mesons also serves the
purpose \cite{GCF}. In this work we propose that the search of
the $B\to D^{*}_{sJ}M$ decays, $M=D$, $\pi$ and $K$, can
discriminate the different theoretical postulations for the
$D^{*}_{sJ}$ content. In the quark-antiquark picture the $B\to
D^{*}_{sJ}M$ branching ratios are expected to be of the same order
of magnitude as the $B\to D^{(*)}_{s}M$ ones, since the
$D^{*}_{sJ}$ meson decay constants should be close to those of the
conventional $D^{(*)}_{s}$ mesons as required by chiral
symmetry \cite{BEH}. We shall assume that the chiral symmetry
is a good symmetry in our analysis. In the unconventional picture the
corresponding decay amplitudes involve additional hard scattering
the four valence quarks of the $D^{*}_{sJ}$ mesons participate.
The branching ratios are then at least suppressed by the coupling
constant and by inverse powers of heavy meson masses, such that
they are smaller than the $B\to D^{(*)}_{s}M$ ones by a factor of
10.

The $B_{d}(P_1)\to D^{*+}_{sJ}(P_2) \pi^-(P_3)$ decay occurs
through the effective Hamiltonian,
\begin{eqnarray}
H_{{\rm eff}}&=&\frac{G_{F}}{\sqrt{2}}V_{ub}V_{cs}^*\left[
C_{1}(\mu ){\cal O}_{1}(\mu)+C_{2}(\mu ){\cal
O}_{2}(\mu)\right]\;, \label{eff}
\end{eqnarray}
with the four-fermion operators ${\cal O}_{1}= (\bar{s}_i
c_j)(\bar{u}_j b_i)$ and ${\cal O}_{2}=(\bar{s}_i c_i)(\bar{u}_j
b_j)$, $(\bar{q}_i q_j)\equiv \bar{q}_i \gamma_{\mu}
(1-\gamma_{5}) q_j$ and $i$ and $j$ being the color indices, the
Cabbibo-Kobayashi-Maskawa (CKM) matrix elements $V$'s, and the
Wilson coefficients $C_{1,2}(\mu )$. We choose a frame, in which
the $B$ meson is at rest and the pion momentum $P_3$ is in the
minus direction in the light-cone coordinates. The two-body decay
rate is expressed as $\Gamma=|A|^2/(16\pi m_B)$, $m_B$ being the
$B$ meson mass and $A$ the decay amplitude.

In the quark-antiquark picture the above decay contains a
color-allowed amplitude, which is written, in the factorization
assumption (FA), as
\begin{eqnarray}
A = i\,\frac{G_F}{\sqrt 2}\,
    V_{ub}^* V_{cs}\,(m_B^2-m_\pi^2)\,f_{D_{sJ}^*}\,
F_0^{B\pi}(m^2_{D_{sJ}^{*}})\,a_1\;, \label{bdfa}
\end{eqnarray}
with the $D_{sJ}^*$ meson decay constant $f_{D_{sJ}^*}$, the
$D_{sJ}^*$ meson (pion) mass $m_{D_{sJ}^{*}}$ ($m_\pi$), and
$a_{1}=C_{2}+C_{1}/N_{c}$, $N_{c}=3$ being the number of colors.
Employing the inputs $G_F=1.16639\times 10^{-5}$ GeV$^{-2}$,
$|V_{ub}|=0.003$, $|V_{cs}|=0.976$, $m_B = 5.28$ GeV,
$\tau_{B^0}=1.542\times 10^{-12}$s, $m_{D_{sJ}^*}=2.32$ GeV, and
$f_{D_{sJ}^*}=0.24$ GeV, $F_0^{B\pi}(m^2_{D_{sJ}^{*}})=0.33$ from
the light-cone-sum-rule results \cite{PB}, and $a_1=1.1$ for a
wide range of the renormalization scale $\mu$, we have the
branching ratio,
\begin{eqnarray}
B(B_{d} \to D^{*+}_{sJ} \pi^-)=3.0\times 10^{-5}\;, \label{fadp}
\end{eqnarray}
close to the Belle and BaBar measurements \cite{Belle,Babar},
$B(B_{d}\to D^{+}_{s} \pi^-)=(2.4^{+1.0}_{-0.8}\pm 0.7,\,
4.6^{+1.2}_{-1.1}\pm1.3)\times 10^{-5}$. Because of
$m_{D_{sJ}^*(2317)}\approx m_{D_{sJ}^*(2463)}$, the result in
Eq.~(\ref{fadp}) holds for both the $D_{sJ}^*(2317)$ and
$D_{sJ}^*(2463)$ mesons.

\begin{figure}[h]
\includegraphics*[width=3.8
in]{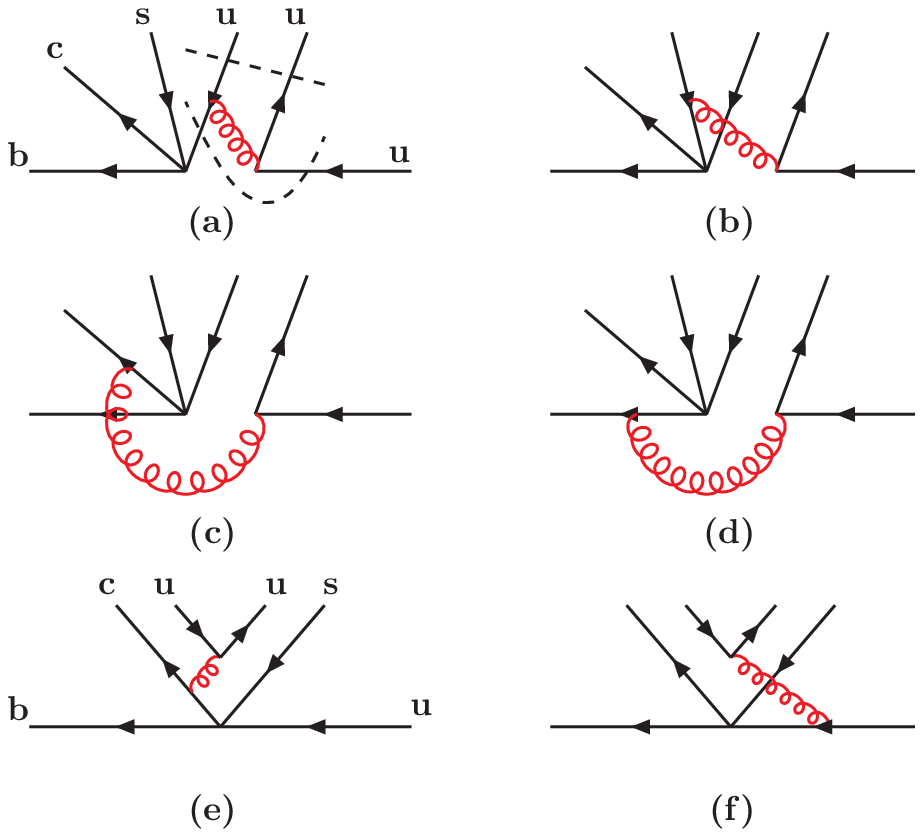}
\caption{(a)-(d) Diagrams contributing to the $B_{d}\to
D^{*+}_{sJ} \pi^-$ decay in the four-quark picture for the
$D^{*}_{sJ}$ content. (e) and (f) Diagrams contributing in the
quark-antiquark picture.} \label{four}
\end{figure}

If the $D^{*}_{sJ}$ meson is a four-quark bound state, the $B_{d}$
meson decays into $D^{*+}_{sJ} \pi^-$ through the diagrams
Figs.~\ref{four}(a)-\ref{four}(d), in which all its four valence
quarks participate hard scattering. An extra hard gluon is then
necessary for producing the $u$-$\bar u$ quark pair, and more
virtual lines appear. For the type of Figs.~\ref{four}(e) and
\ref{four}(f), the exchanged gluon, being of collinear origin with
the momentum in the plus direction, should be absorbed into the
two-parton $D^{*+}_{sJ}$ meson distribution amplitude. That is,
Figs.~\ref{four}(e) and \ref{four}(f) contribute to the analysis
in the quark-antiquark picture.

There are two color configurations,
\begin{eqnarray}
\frac{1}{N_c^2}c^b{\bar s}^b n^c {\bar n}^c\;,\;\;\;
\frac{1}{12}f^{ab_1c_1}f^{ab_2c_2}c^{b_1}{\bar s}^{c_1} n^{b_2}
{\bar n}^{c_2}\;,
\end{eqnarray}
where both the $c\bar s$ and $n\bar n\equiv(u\bar u+d\bar
d)/\sqrt{2}$ pairs are in the color-singlet 1 and color-octet 8
states, respectively \cite{Zh}. The average over colors has been
made explicit. The Wilson coefficients associated with each
diagram from the 11 and 88 configurations are listed in
Table~\ref{wc1}. It is found that the Wilson coefficient for
Fig.~\ref{four}(a) from 11 is largest. The contributions from
Figs.~\ref{four}(b) and \ref{four}(c), besides a pair cancellation
\cite{KKL}, are down by a small ratio $C_1/a_1\sim -0.2$. As shown
later, the amplitude corresponding to Fig.~\ref{four}(d), where
the hard gluon attaches the $\bar b$ quark, is suppressed by a
power of $\Lambda_{\rm QCD}/m_{D^*_{sJ}}\sim 0.1$, though it is
not down by a Wilson coefficient. Hence, we can safely drop
Figs.~\ref{four}(b)-\ref{four}(f), and consider only
Fig.~\ref{four}(a) from the 11 color configuration.

\begin{table}[htb]
\caption{Wilson coefficients associated with the diagrams in
Fig.~1.} \label{wc1}
\begin{ruledtabular}
\begin{tabular}{c|cccc}
configuration & (a) & (b) & (c) & (d) \\\hline
11 & $a_1/N_c$ & $C_1/N_c$ & $C_1/N_c$ & $a_1/N_c$\\
88 & $C_1/N_c$ & $C_1/N_c$ & $C_1/N_c$ & $C_1(1/N_c-1)$\\ \hline
\end{tabular}
\end{ruledtabular}
\end{table}

A quantitative analysis of Fig.~\ref{four} requires the knowledge
of the four-parton $D^{*}_{sJ}$ meson distribution amplitude.
Before this information is available, we make a simple estimation
also in FA. Insert the Fierz identity,
\begin{eqnarray}
1_{ij}1_{lk}&=&\frac{1}{8}(1-\gamma_5)_{ik}(1-\gamma_5)_{lj}
+\frac{1}{8}(1+\gamma_5)_{ik}(1+\gamma_5)_{lj}
+\frac{1}{8}[\gamma_\nu(1-\gamma_5)]_{ik}
[(1-\gamma_5)\gamma^\nu]_{lj}
\nonumber\\
& &+\frac{1}{8}[\gamma_\nu(1+\gamma_5)]_{ik}
[(1+\gamma_5)\gamma^\nu]_{lj}
+\frac{1}{8}(\sigma^{\nu\lambda})_{ik}(\sigma_{\nu\lambda})_{lj}\;,
\label{fi1}
\end{eqnarray}
into Fig.~\ref{four}(a) to factorize the fermion flows. The first
term, inserted in the way indicated by the lower dashed line,
gives the factorization of the $B\to\pi$ form factor from the full
amplitude. The insertion of the third term indicated by the upper
dashed line then leads to a nonvanishing hard kernel and to the
matrix element $\langle D^{*+}_{sJ}|\bar{c}\gamma_\mu(1-\gamma_5)
s \bar{u}\gamma_\nu(1-\gamma_5) u|0\rangle$, which defines the
$D^{*}_{sJ}$ meson decay constant. There exists another
factorization of fermion flows with the fourth (first) term in
Eq.~({\ref{fi1}) inserted at the lower (upper) dashed line.
However, this factorization introduces the matrix element $\langle
D^{*+}_{sJ}|\bar{c}\gamma_\mu(1-\gamma_5) s \bar{u}(1-\gamma_5)
u|0\rangle$, which is suppressed by a power of $m_{D_{sJ}^*}/m_B$
compared to the previous one.

We derive the $B_{d}\to D^{*+}_{sJ} \pi^-$ decay amplitude in FA
in the four-quark picture,
\begin{eqnarray}
A = \frac{G_F}{\sqrt 2}\, V_{ub}^* V_{cs}\,
    \langle D^{*+}_{sJ}|\bar{c}\gamma_\mu(1-\gamma_5)
s \bar{u}\gamma_\nu(1-\gamma_5) u|0\rangle\,
\langle\pi^-|\bar{b}\gamma^\mu(1-\gamma_5) u|B_{d} \rangle
\,a_1\; H^\nu\;,\label{bdfs}
\end{eqnarray}
with the hard kernel,
\begin{eqnarray}
H^\nu=\frac{g^2}{32\sqrt{2}}\frac{C_F}{N_c}\frac{tr[
(1-\gamma_5)\not l_u\gamma_\beta
(1-\gamma_5)\gamma^\nu\gamma^\beta]}{l_u^2 l_g^2}\;, \label{ha0}
\end{eqnarray}
where $l_u$ and $l_g$ are the momenta carried by the internal $u$
quark and gluon, respectively, and the denominator $\sqrt{2}$
comes from the definition of $n\bar n$. To be precise, $H^\nu$
should be expressed as a convolution of Eq.~(\ref{ha0}) with the
four-parton distribution amplitude over the momentum fractions of
the valence $\bar s$, $u$ and $\bar u$ quarks. For the purpose of
estimation, we regard that these valence quarks carry the fixed
momentum fractions of $O(\Lambda_{\rm QCD}/m_{D^*_{sJ}})$
\cite{TLS2}. Therefore, the components of $l_u$ and $l_g$ have the
orders of magnitude,
\begin{eqnarray}
l_u\sim l_g\sim \frac{m_B}{\sqrt{2}} \left(\frac{\Lambda_{\rm
QCD}}{m_{D^*_{sJ}}}, \frac{1}{2}, {\bf 0}_T\right)\;, \label{lug}
\end{eqnarray}
where the valence $\bar u$ quark in the pion has been assumed to
take half of the pion momentum. The virtual $\bar b$ quark
momentum in Fig.~1(d) has the components $l_b\sim
(m_B/\sqrt{2})(1, 1/2, {\bf 0}_T)$, such that Fig.~\ref{four}(d)
is power-suppressed by $l_u^2/l_b^2\sim \Lambda_{\rm
QCD}/m_{D^*_{sJ}}$ compared to Fig.~\ref{four}(a) as stated
before.

Next step is to evaluate the matrix element,
\begin{eqnarray}
\langle D^{*+}_{sJ}(P_2)|\bar{c}\gamma_\mu(1-\gamma_5) s
\bar{u}\gamma_\nu(1-\gamma_5) u|0\rangle=i\frac{B}{m_{D^*_{sJ}}}
P_{2\mu} P_{2\nu} f_{D^*_{sJ}}\;,
\end{eqnarray}
which has been parametrized in terms of a dimensional constant
$B$. Under the heavy quark symmetry, this matrix element should be
close to $\langle D^{0}|\bar{c}\gamma_\mu(1-\gamma_5) u
\bar{u}\gamma_\nu(1-\gamma_5) u|0\rangle$. The equation of motion
for the heavy $c$ quark with the momentum $P_c\approx P_2$ and the
relation $P_2^2=m_{D}^2$ lead to $\langle
D^{0}(P_2)|\bar{c}(1-\gamma_5) u \bar{u}\gamma_\nu(1-\gamma_5)
u|0\rangle=iBP_{2\nu} f_{D}$ with the $D^{0}$ meson decay constant
$f_{D}$. The Fierz transformation of the four-quark operator and
FA of the matrix element give $\langle
D^{0}(P_2)|\bar{c}\gamma_\nu (1-\gamma_5) u \bar{u}(1-\gamma_5)
u|0\rangle\approx \langle
D^{0}(P_2)|\bar{c}\gamma_\nu(1-\gamma_5)u|0\rangle \langle 0|
\bar{u}(1-\gamma_5) u|0\rangle$. Substituting the definition of
$f_{D}$, we derive
\begin{eqnarray}
B\approx \langle 0|\bar{u}(1-\gamma_5) u|0\rangle =\langle 0|
\bar{u} u|0\rangle\approx -0.24\;\;{\rm GeV}^3\;,
\end{eqnarray}
where the standard value of the quark condensate has been adopted.

Equation (\ref{bdfs}) then becomes
\begin{eqnarray}
A = i\frac{G_F}{\sqrt 2}\, V_{ub}^* V_{cs}\,(m_B^2-m_\pi^2)\,f_{D_{sJ}^*}\,
    F_0^{B\pi}(m^2_{D_{sJ}^{*}})\,a_1\, R\;, \label{bd1}
\end{eqnarray}
with the ratio,
\begin{eqnarray}
R&=&\frac{g^2C_F}{32\sqrt{2}N_c}B\frac{tr[(1-\gamma_5)\not
l_u\gamma_\beta (1-\gamma_5)\not
P_2\gamma^\beta]}{m_{D^*_{sJ}}l_u^2 l_g^2}\;,
\nonumber\\
&=&-\sqrt{2}\pi^2\left(\frac{\alpha_s}{\pi}\right)
\frac{C_F}{N_c}\frac{B}{m_{D^*_{sJ}}\Lambda_{\rm QCD}^2}
\left(\frac{m_{D^*_{sJ}}}{m_B}\right)^2 \approx 0.275\;,
\label{ha2}
\end{eqnarray}
for the inputs $\alpha_s/\pi=0.2$ in $b\to c$ transitions
\cite{TLS2} and $\Lambda_{\rm QCD}\approx 0.3$ GeV. $l_u^2$ and
$l_g^2$ from Eq.~(\ref{lug}) have been inserted. It is easy to see
that the decay amplitude in the four-quark picture is down by the
coupling constant $\alpha_s$, by the color number $1/N_c$, and by
the powers $(m_{D^*_{sJ}}/m_B)^2$.
We conclude that the $B_{d}\to D^{*+}_{sJ}\pi^-$ branching ratio
in the four-quark picture should be smaller than that in the
quark-antiquark one by a suppression factor,
\begin{eqnarray}
\frac{B^{(4)}(B_{d}\to D^{*+}_{sJ} \pi^-)} {B^{(2)}(B_{d}\to
D^{*+}_{sJ} \pi^-)}=R^2\approx 0.08\;.
\end{eqnarray}
If the $B_{d}\to D^{*+}_{sJ} \pi^-$ branching ratios are observed
at the $10^{-5}$ level as in Eq.~(\ref{fadp}), the $D^{*+}_{sJ}$
meson is likely to be a conventional quark-antiquark state. If it
is observed with the $10^{-6}$ (around $2.4\times 10^{-6}$)
branching ratio, the four-quark picture is preferred.

There is already a hint from the $B\to D^{*}_{sJ}D$ decays, to
which our analysis can be generalized straightforwardly simply by
substituting the $B\to D$ form factor for the $B\to\pi$ form
factor. The $B\to D^{*}_{sJ}D$ branching ratios have been measured
by Belle recently \cite{BeF}:
\begin{eqnarray}
& &B(B^+\to D^{*+}_{sJ}(2317)\bar D^0)\times B(D^{*+}_{sJ}(2317)\to
D^+_s\pi^0)=(8.1^{+3.0}_{-2.7}\pm 2.4) \times 10^{-4}\;,
\nonumber\\
& &B(B^+\to D^{*+}_{sJ}(2463)\bar D^0)\times B(D^{*+}_{sJ}(2463)\to
D^{*+}_s\pi^0) =(11.9^{+6.1}_{-4.9}\pm 3.6)\times 10^{-4}\;,
\nonumber\\
& &B(B^+\to D^{*+}_{sJ}(2463)\bar D^0)\times B(D^{*+}_{sJ}(2463)\to
D^+_s\gamma) =(5.6^{+1.6}_{-1.5}\pm 1.7)\times 10^{-4}\;.
\end{eqnarray}
The first data together with
$B(B^+\to D^+_{s}\bar D^0)= (1.3\pm 0.4)\%$ \cite{PDG} imply
\begin{eqnarray}
\frac{B(B^+\to D^{*+}_{sJ}(2317)\bar D^0)}
{B(B^+\to D^+_{s}\bar D^0)}\approx 0.06\;,
\label{rbr}
\end{eqnarray}
and the four-quark content of the $D^{*}_{sJ}$ meson. The latter
two data, assuming that the $D^{*}_{sJ}(2463)$ decays only through
the channels $D^*_s\pi^0$ and $D_s\gamma$, lead to $B(B^+\to
D^{*+}_{sJ}(2463)\bar D^0)\approx 0.18\%$ and the ratio,
\begin{eqnarray}
\frac{B(B^+\to D^{*+}_{sJ}(2463)\bar D^0)}
{B(B^+\to D^+_{s}\bar D^0)}\approx 0.14\;,
\label{rbr1}
\end{eqnarray}
which also gives a similar indication. It is unlikely that
the dramatically different branching ratios in Eq.~(\ref{rbr}) is
due to the different decay constants $f_{D_{sJ}^*}$ and
$f_{D_{s}}$ \cite{DO} from the viewpoint of heavy quark symmetry.


At last, we discuss another ideal mode for the purpose, the
$B_{d}\to D^{*-}_{sJ} K^{(*)+}$ decay, which occurs through the
operators ${\cal O}_{1}= (\bar{d}_i u_j) (\bar{c}_j b_i)$ and
${\cal O}_{2}= (\bar{d}_i u_i) (\bar{c}_j b_j)$ with the product
of the CKM matrix element $V_{cb}V_{ud}^{*}$. Since this mode
involves only the annihilation topologies, FA does not apply.
Hence, we estimate its branching ratio in the quark-antiquark
picture using the perturbative QCD (PQCD) approach, in which a
transition matrix element is expressed as the convolution of hard
kernels of the valence quarks with hadron distribution amplitudes
\cite{LB,KLS}.
The derivation of the factorization formulas at leading power in
$1/m_B$ and leading order in $\alpha_s$ follow that for the $B\to
D^{(*)}\pi(\rho,\omega)$ decays in \cite{KKL}. We shall present
the explicit expressions elsewhere. Adopting the $D^*_{sJ}$ meson
distribution amplitudes the same as those for the $D^{(*)}$ meson
in \cite{KKL} (the $B$, $K$ and $K^*$ meson distribution
amplitudes have been known from the literature), we obtain the
branching ratios listed in Table~\ref{cd}. The predictions for the
$B_{d}\to D^{*-}_{sJ} K^{+}$ mode are close the Belle and BaBar
measurements \cite{Belle,Babar}, $B(B_{d}\to D^{-}_{s}
K^{+})=(3.2\pm 0.9 \pm 1.0,\, 3.2 \pm 1.0 \pm 1.0)\times 10^{-5}$,
which are in agreement with the PQCD predictions
\cite{Kurimoto,Chen-PLB}. The estimation for the $B_{d}\to
D^{*-}_{sJ} K^{(*)+}$ branching ratios in the four-quark picture
is similar, and the results are also smaller than those in the
quark-antiquark picture by a factor about 0.08. For example, the
$B_{d}\to D^{*-}_{sJ}(2317) K^{+}$ branching ratio is expected to
be around $4.2\times 10^{-6}$.

Our estimation given above applies to other
non-quark-antiquark models for the $D^{*}_{sJ}$ content, such as a
molecule, up to order of magnitude. One of the differences is that
the 88 color configuration is excluded, which is not essential
anyway.

\begin{table}[htb]
\caption{$B_{d}\to D^{*-}_{sJ} K^{(*)+}$ branching ratios (in
units of $10^{-5}$) in the quark-antiquark picture from the PQCD
approach.} \label{cd}
\begin{ruledtabular}
\begin{tabular}{ccc}
$B_{d}\to D^{*-}_{sJ}(2317) K^{+}$ & $B_{d}\to D^{*-}_{sJ}(2317)
K^{*+}$ & $B_{d}\to D^{*-}_{sJ}(2463) K^{+}$
\\\hline
5.35 & 7.79 & 7.01
\end{tabular}
\end{ruledtabular}
\end{table}


In summary, a measurement of the $B\to D^{*}_{sJ}M$ branching
ratios, $M=D$, $\pi$ and $K$, can provide more information on the
nature of the new $D^{*}_{sJ}$ mesons. If the $D^{*}_{sJ}$ mesons
are multi-quark bound states, they will be more difficult to be
produced than the conventional $D_s^{(*)}$ mesons in exclusive $B$
meson decays: the branching ratios will be one order of magnitude
smaller. The suppression is a combined effect of $\alpha_s$,
$1/N_c$ and $(m_{D^*_{sJ}}/m_B)^2$, which arise from the
additional hard scattering the four valence quarks of the
$D^{*}_{sJ}$ mesons participate. More precise data are necessary
for drawing a conclusion, though the recently measured
$B\to D^{*}_{sJ}D$ branching ratios have indicated an
unconventional picture.

\vskip 0.5cm

The authors thank  H.Y. Cheng, W.S. Hou and Taekoon Lee for useful
discussions. This work was supported in part by the National
Science Council of R.O.C. under Grant No. NSC-91-2112-M-001-053
and by the National Center for Theoretical Sciences of R.O.C..

\end{document}